\documentclass{article}

\usepackage[dvips]{graphicx}

\title{The Challenge of Editing\\ Einstein's Scientific Manuscripts}

\author{Tilman Sauer\thanks{Paper presented at the 25th Anniversary
Annual Meeting of the Association for Documentary Editing, Chicago,
November 14--16, 2003; to appear in {\it Documentary Editing}. 
Facsimiles of Einstein's manuscripts are reproduced by kind permission 
of the Albert Einstein Archives, Jerusalem.}}  

\date{\today}

\begin{document}

\maketitle

\begin{abstract}

Einstein's research manuscripts provide important insights into his
exceptional creativity. At the same time, they can present
difficulties for a publication in the documentary edition of the {\it
Collected Papers of Albert Einstein} (CPAE). The problems are
illustrated by discussing how some important examples of Einstein's
research manuscripts have been included in previous volumes of the
CPAE series: his Scratch Notebook from the years 1910--1914, his
so-called Zurich Notebook from 1912, documenting his early search for
a generally covariant theory of gravitation, and the
Einstein-Besso manuscript from 1913, containing calculations of
Mercury's perihelion advance on the basis of the Einstein-Grossmann
equations. Another category of research notes are small and
disconnected ``back-of-an-envelope'' calculations. A major challenge
for future volumes of the CPAE series are Einstein's Berlin and
Princeton research manuscripts on a unified field theory. This batch
of some 1700 undated manuscript pages presents a formidable challenge
also for historians of science. Although the web provides new
possibilities for the editorial task, such as the publication of
facsimiles on the {\it Einstein Archives Online} website, it is argued
that a satisfactory solution of the editorial problems posed by these
manuscripts depends on scholarly efforts to reconstruct and understand
the content of Einstein's manuscripts.

\end{abstract}

\section{The Einstein Papers Project}

The {\it Einstein Papers Project} is a long-term editorial project
devoted to publishing the {\it Collected Papers of Albert Einstein}
(CPAE).\footnote{For general information about the Einstein Papers
Project, see the ``Publisher's Foreword'' and the ``General
Introduction'' in Volume 1 of the series \cite{CPAE1}, as well as
\cite{Bailey1989,Stachel1987,Buchwald2004}. The project's website is
at www.einstein.caltech.edu.} The first volume was published by
Princeton University Press in 1987 \cite{CPAE1}, followed by eight
more volumes to date \cite{CPAE2}--\cite{CPAE9}. To complete the
series, some twenty more volumes are anticipated during the next
30--40 years. The documentary edition of the CPAE is supplemented by
an English translation series. In addition to these publications in
book format, the project has launched a website jointly with the
Albert Einstein Archives of the Hebrew University of Jerusalem. Known
as {\it Einstein Archives Online} (www.alberteinstein.info), the site
presently provides over 3,000~high-quality facsimiles from Einstein's
autograph writings, a finding-aid to the Albert Einstein Archives, and
an itemized database of some 43,000~records.

The editorial project draws on the collection of the Albert Einstein
Archives. The core of the archives was put together in Princeton after
Einstein's death by his long-time secretary Helen Dukas who acted as
archivist in a devoted effort until her death in 1982. The ``Dukas
collection'' comprises some 43,000 documents. The collection was
microfilmed into 61 reels during the seventies,\footnote{Hence
archival call numbers for items of the Einstein Archives are of a
two-number format, e.g.\ 3-006, where the first number indicates the
reel and the second number the sequentially numbered document in that
reel.}  and a hardcopy duplicate archive was produced from the
microfilm and collated with the originals for editorial
purposes. Following Einstein's will which bequeathed his literary
estate to the Hebrew University of Jerusalem, the collection was
shipped to Jerusalem after Helen Dukas's death, and has been housed
since then at the Jewish National and University Library. Since 1982
some 20,000 additional documents have been collected both by curators
of the Albert Einstein Archives and by editors of the CPAE. These
documents, mostly hardcopies from archives all over the world, were
added to the original collection in a supplementary archive.

The documentary edition of the CPAE publishes all of Einstein's
scientific writings, both published and unpublished, as well as
drafts, notebooks, scientific and personal correspondence, in
chronological order. With the exception of the first volume, covering
the early years 1879--1902, the published volumes have been divided
into a {\it Writings} series \cite{CPAE2,CPAE3,CPAE4,CPAE6,CPAE7} and
a {\it Correspondence} series \cite{CPAE5,CPAE8,CPAE9}. The editorial
method follows rigid standards of documentary editing. In the Writings
volumes, published items are reproduced in facsimile; comparisons to
drafts or versions are examined and detailed in the
footnotes. Unpublished materials are transcribed, maintaining
substantial faithfulness to the original text. No silent corrections
of typographical or other errors are applied, and punctuation and
style are reproduced, while errors of fact or calculation are
explained in the annotation.\footnote{A detailed account of the
editorial method is given in toto in \cite{CPAE1} with volume-specific
supplements in subsequent volumes. Beginning with \cite{CPAE8} the
full editorial method is reproduced in each volume.} Specific
references to persons, places, literature, scientific developments,
organizations, and events are identified. Details of the annotation
follow the general aim of ``bridging the gap'' of knowledge familiar
to the author or intended audience and that of contemporary
readers.  An introduction to each volume and various editorial
headnotes analyze major themes in Einstein's life and work.

The website {\it Einstein Archives Online} presently publishes
high-quality facsimiles of all of Einstein's autograph writings extant in
the original Dukas collection. Presently, no correspondence is
included, nor does the website present typescripts or third-party
documents. Eight important autographs whose facsimiles are provided on
the site but that are not part of the AEA belong to the Schwadron and
Yahuda Collections held at the Jewish National and University
Library. The website also presents PDF versions of those 39 documents
in the published volumes of the CPAE for which autographs are also
presented in facsimile.

In addition to the original documents, the website provides a
finding-aid to the original Einstein Archives. The finding-aid
provided on the website is an HTML transform using XSL style sheets
from an XML file incorporating the Encoded Archival Description (EAD)
markup.\footnote{The EAD Document Type Definition (DTD) is a standard
for encoding archival finding aids using the Standard Generalized
Markup Language (SGML) and is maintained in the Network Development
and MARC Standards Office of the Library of Congress, in partnership
with the Society of American Archivists.}

The itemized archival database that is made accessible on the website
is a subset of a master archival database that is jointly maintained
by both the Albert Einstein Archives and the Einstein Papers
Project. It provides basic information on author, receiver, dating,
title, language, location, and physical description of a document. If
applicable, the database also provides publication information in the
CPAE. The archival database covers all items of the original Dukas
collection plus those items of the ``supplementary archives'' that
have already been published in the CPAE series (634 records). The
archival database information displayed on the website is intended to
facilitate general access to the holdings of the Einstein Archives. In
contrast to the editorial apparatus in the documentary edition, where
every effort is made to check the accuracy of any information against
the original sources, no guarantuee is given as to the accuracy,
consistency, or completeness of the information displayed in the
database records. In fact, the database records are continuously
revised and periodically updated as additional research is carried out
at both the Albert Einstein Archives and the Einstein Papers Project.

Neither the {\it Collected Papers of Albert Einstein} nor the {\it
Einstein Archives Online} is concerned with non-textual material in
the strict sense. The Einstein Papers Project does not edit any audio
material unless as transcripts, nor does it include pictures or
images, other than select images for illustrative purposes. There are
no concerns about material witnesses like experimental setups,
astrographical plates, or the like. There is, however, a category of
textual material that is of central importance for the edition and
that defies to some extent standard procedures of documentary editing,
i.e.\ Einstein's research manuscripts.

\section{Einstein's Scientific Manuscripts}

Albert Einstein (1879--1955) is a household name, a synonym for
scientific creativity and responsibility. He was honored as ``person
of the $20^{\rm th}$ century'' by {\it TIME} magazine. In 2005, all over
the world numerous conferences, lectures, exhibitions, and
other events are being planned for the centennial anniversary of his
``miracle year.'' It was in 1905 when Einstein, an unknown technical
expert at the Swiss Federal patent office in Bern, published, within a
few months, a series of five papers, each of which had a profound and
lasting impact on the development of twentieth century
physics.\footnote{The papers are reprinted as facsimiles with
extensive editorial annotation as Docs.~14, 15, 16, 23, and 24 in
\cite{CPAE2}. As an off-shoot of the editorial project the papers are
also available, in English translation, as \cite{Stachel1998}.} The
papers deal with the determination of molecular dimensions, give an
explanation of the phenomenon of Brownian motion, expound what we now
call the Special Theory of Relativity, including the equivalence of
mass and energy captured in the famous equation $E=mc^2$, and present
an explanation of the photoelectric effect by putting forward the
light quantum hypothesis. The latter contribution alone later earned
Einstein the Nobel Prize for 1921. 

Ten years later, Einstein had risen up through the ranks of academic
hierarchy as {\it Privatdozent} in Bern, Extraordinary and Full
Professor in Zurich and Prague, and in 1914 had accepted a position
without teaching obligations as member of the Prussian Academy of
Science in Berlin. It was then that Einstein crowned his earlier
scientific contributions with another conceptual breakthrough that came after
years of strenuous efforts to generalize the special theory of
relativity to include gravitation. The General Theory of Relativity
was completed with the publication of generally covariant field
equations of gravitation in late 1915. It is these equations that even
today are the basis for extensive research, both theoretical and
experimental, e.g. with respect to finding evidence for gravitational
waves.

In addition to his major contributions of 1905 and 1915, Einstein
published numerous papers of significant importance in various fields
of physics. These papers include theoretical investigations into the
foundations of kinetic theory, statistical physics and radiation
theory, work on the law of photochemical equivalence, on the specific
heat of solids at low temperatures, on the phenomenon of opalescence,
on the so-called Einstein-De Haas experiment to determine the
relationship between magnetic moments and moments of inertia, a number
of critical investigations into the foundations of quantum physics,
and some ground-breaking investigations into the consequences of
general relativity, such as gravitational waves, cosmological
consequences, equations of motion, and gravitational lensing. During his
later years, Einstein worked intensely on the problem of finding a
unified field theory of gravitation and electromagnetism that would
also account for the structure of matter and the quantum phenomena.

Given Einstein's exceptional significance as a highly creative,
successful, and prodcutive scientist, a natural and significant
interest for an edition of his works arises from the wish to better
understand the working and circumstances of his productivity. His
burst of creativity in his miracle year still provides a major
challenge for a historical reconstruction that would in some sense
explain his creativity of this year. Unfortunately any
such attempts are hampered by the scarcity of documentary evidence
even after publication of the respective volumes of the CPAE series
\cite{CPAE1,CPAE2,CPAE5}. The situation is much better for Einstein's
search for a General Theory of Relativity where much more pertinent
documents have survived and are now available. It is in the context of
the attempt to come to a better historical understanding of Einstein's
thinking that his research manuscripts acquire a special importance.

The term ``research manuscript'' here refers to a document that was
written in the creative process of thinking about a scientific
problem, mainly for the purpose of developing one's own thoughts and
of realizing implicit consequences inherent in a mathematical
formalism. The manuscripts are written without any potential audience
or readership in mind, other than the author himself. If the creative
work involves two individuals that are collaborating, the manuscript
may serve for communication between the researchers. A common
characteristic of research manuscripts in Einstein's case is an
abundance of mathematical formulae without, or with only very few,
explicit words. In most cases, the manuscripts are not dated, neither
explicitly nor indirectly. In many cases, the calculations are on
single sheets of paper but in a few cases they are contained in bound
notebooks.

Research manuscripts are the most immediate written evidence of a
creative process since they are produced without further reflection
during that process.\footnote{For a general perspective on the
significance of research notes and notebooks for the history of
science, see \cite{Holmesetal2003}.} Later stages of writing about a
scientific topic tend to be more organized or organized differently,
and to the extent that the composer of the manuscript is aware of his
addressing a potential audience, he begins to explicate the tokens of
the abstract formalism and tries to convey the meaning of the
mathematical expressions by linking it with words of natural or
scientific language.

Given the significance of the creative process in Einstein's case, his
research manuscripts carry a unique value for historians of
science. At the same time, they often pose a problem for the explicit
rules of editorial procedure laid out in the editorial method. These
problems pertain most importantly to the uncertainty of dating and to
the uncertainty of the coherence and sequence of calculations.

In the following I will discuss four examples of research manuscripts
that have already been published in the {\it Collected Papers of
Albert Einstein} and one example of a batch of research manuscripts
that poses a formidable problem for future volumes of the series.

\subsection{Einstein's Scratch Notebook, 1910--1914?}

The first {\it Writings} volume of the CPAE series covers the years
1902--1909 \cite{CPAE2}. It thus contains his very first published
papers, all of his famous publications from his annus mirabilis
1905, as well as all later publications that he wrote while working at
the Patent Office in Bern. Sadly, no manuscripts from this period are
extant and all documents of this volume are reprinted facsimiles of
his publications.

This situation changes with Volume 3 \cite{CPAE3} which covers the
years 1909--1911 when Einstein accepted his first call as an
Extraordinary Professor at the University of Zurich and, a little
later, another call as Full Professor at the German University of
Prague. This volume reprints some twenty published documents but also
publishes seven manuscripts. Most of these posed no problem for the
editorial method. Three items are lecture notes that Einstein wrote in
preparation of academic courses which he had to give as a professor in
Zurich and Prague or for a lecture that he gave in Zurich in 1911. Two
items are short manuscripts: one was a written response to a paper by
Planck, another one was a statement written in response to a request
by the Berlin autograph collector Darmstaedter. Another item are
handwritten discussion remarks following lectures delivered at the
first Solvay Congress in late 1911. These manuscripts, in each case,
are perfectly coherent and datable documents.

The one research manuscript included in Volume 3 that did pose a
problem to the editors was eventually printed in the appendix. The
editors preface this appendix by the following comments:
\begin{quote}
This notebook was probably purchased by Einstein in 1909 when he began
his appointment at the University of Zurich. It bears a sticker of the
Zurich stationer Landolt and Arbenz. The last entries suggest that
Einstein did not use the notebook after taking up a position in Berlin
in 1914. The disparate nature and discontinuity of entries
(e.g., diagrams, equations, notes on appointments, references to
scientific literature, and addresses), as well as their disjointed
chronological sequence, argue for preserving unity in the presentation
of this notebook. It is printed here in its entirety in facsimile,
with accompanying pages of transcription to make it readily
accessible.\cite[p.~563]{CPAE3}
\end{quote}
The notebook is then presented with facing pages in such a way that
the top half of each page of the volume presents the facsimile of two
facing pages of the Scratch Notebook. Below the facsimile, on the
bottom half of each page, a conformal transcription of the page is
given. Following the notebook, the editors provide a descriptive note
and a list of the literature referenced in the notebook, in the order
of appearance.

The context of a few significant pages of this scratch notebook that
clearly contain research calculations was identified some time after
its publication in the CPAE \cite{Rennetal1997,RennSauer2003a}. It was
shown that eight pages of the Scratch Notebook contain calculations
that are fully equivalent to calculations that were only published by
Einstein in 1936---some 24 years later---in a paper on what is now 
known as gravitational lensing (see Fig.~1).
\begin{figure}
\begin{center}
\includegraphics[scale=.6]{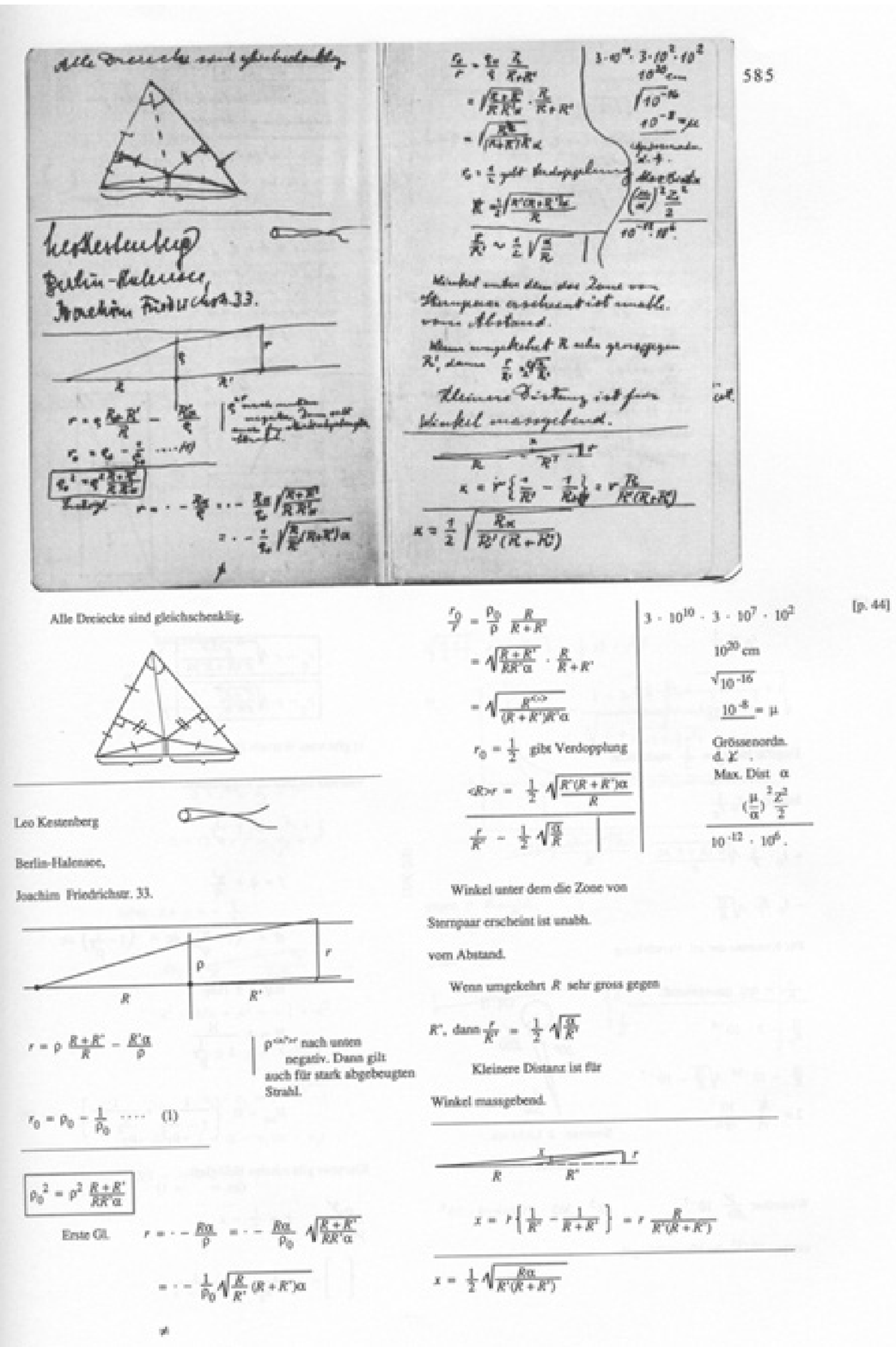}
\caption{{\it Einstein's Scratch Notebook}, 1910--1914?, was reproduced as
facsimile with accompanying conformal transcription in
\cite[App.~A]{CPAE3}. This page \cite[p.~585]{CPAE3} shows
Einstein's earliest calculations about gravitational lensing. For a
high-quality facsimile, see Doc.~3-013, images 23 and 24, of {\it
Einstein Archives Online} at www.alberteinstein.info. \copyright~The
Hebrew University of Jerusalem, Albert Einstein Archives. }
\end{center}
\end{figure}

The differences amount to mere differences of mathematical notation,
and the pages could be dated, by reference to neighboring pages and
historical context, to a week in April 1912. The reconstruction also
confirmed a feature that was evident to readers of the documentary
edition only through the descriptive note. One page of the notebook
was a loose page inserted within the bound pages and was facsimilized
together with the pages where it was physically found. It nevertheless
contained a part of the calculation that was independent from the flow
of the calculation in the bound part of the notebook. The phenomenon
of gravitational lensing was first observationally confirmed in 1979
and today presents a highly active field of modern astrophysical
research.

\subsection{Research Notes on a Generalized Theory of Relativity}

Volume 4 of the CPAE series covers the writings of the years
1912--1914, up until Einstein's move to Berlin in April 1914
\cite{CPAE4}. It is a period in which Einstein is deeply involved in a
search for a generalized theory of relativity and a theory of
gravitation. This search produced two research manuscripts that posed
new problems for the editorial method.  One is a bound notebook with
research notes by Einstein, dealing mainly with the theory of
gravitation. It is commonly referred to as the ``Zurich Notebook''
because it dates from Einstein's time as a professor at the ETH
Zurich. It documents Einstein's search for a gravitational field
equation from the first insight into the significance of the metric
tensor in summer 1912 until he settled on a preliminary theory of
gravitation that he published together with his friend and
colleague Marcel Grossmann in an {\it Entwurf}, i.e.\ ``Outline'' of
the theory of spring 1913 (see \cite[Doc.~13]{CPAE4}). The other one
is a research manuscript consisting of loose sheets, written partly by
Einstein, partly by his friend and collaborator, Michele Besso, and
deals with the problem of calculating the motion of the perihelion of
Mercury on the basis of the preliminary {\it Entwurf} theory of
gravitation.

The significance of the Zurich Notebook for a detailed
understanding of Einstein's path towards a general theory of
relativity was first realized by John Stachel in the course of
preparing the editorial project of the CPAE. It was then studied by
John Norton who based some decisive arguments in his ground-breaking
1984 account of Einstein's search for his gravitational field
equations on a reconstruction of some pages of this document
\cite{Norton1984}. While the content of a major portion of this
notebook was thus clearly identified as dealing with the problem of
gravitation and could be related on the basis of content and
idiosyncratic notation quite unambiguously to Einstein's publications
between summer 1912 and spring 1913, many problems remained. Norton
had based his 1984 account only on a reconstruction of a few
individual pages of the notebook, and although some pages in the later
part of the notebook could be reconstructed as coherent calculations
that extended over several pages, the majority of the entries were
still unidentified. Two other problems arose. First, entries were made
in the bound notebook starting at two ends, with one page showing
entries from both sides where the flows of entries met. This fact as
well as indications of later corrections made a determination of the
sequence of entries ambiguous despite the fact that the pages have a
natural sequence in the bound notebook. Second, about a third of the
entries in the notebook were apparently not dealing with gravitation
but with other problems of statistical physics and radiation
theory. Despite efforts by the editors to reconstruct the
meaning of these parts of the notebook, they remained unidentified.

The editors of Volume 4 of the series decided to publish only that
portion of the ``Zurich Notebook'' that deals with gravitation, and
only mentioned and briefly described the other part of the notebook in
the descriptive note \cite[Doc.~10]{CPAE4}. The part that deals with
gravitation theory was presented in conformal transcription. In order
to facilitate access to the manuscript, the document was prefaced by an
editorial note that elaborates on the context and content of the
research notes. In a slight deviation of the general rules of
presentation, explanatory notes to the text were printed as footnotes
rather than endnotes. The annotation is sparce but does provide some
information pertaining to a reconstruction of the calculations,
reflecting the degree of understanding obtained by the editors at that
time.

The significance of the Zurich Notebook for the history of science
lies in the fact that it allows for unique insights into a crucial
period of Einstein's path towards a general theory of relativity
\cite{RennSauer2003b}. The attention that the notebook received in the
course of preparing the pertinent volume of the {\it Collected Papers}
triggered a research effort of five historians of science, all of them
past or present editors of the project.\footnote{See
\cite{Rennetal2005} and, for a preliminary account of results of
this research effort, \cite{RennSauer1999}.} The group embarked on a
line-by-line reconstruction of the gravitational parts of the
notebook. They found that it documents Einstein's growing familiarity
with an unknown formalism of tensor calculus following his insight
that the metric tensor would have to play a crucial role in the
formulation of the general theory of relativity. The notes, in fact,
document Einstein's search for a gravitational field equation with the
investigation of a sequence of candidate equations that are being
tested against heuristic requirements. Among other things, it turned
out that Einstein investigated the correct field equations of
gravitation already in 1912, if only in linear approximation, but at
that time discarded them as unacceptable from a physical point of view
(see Fig.~2).
\begin{figure}
\begin{center}
\includegraphics[scale=.6]{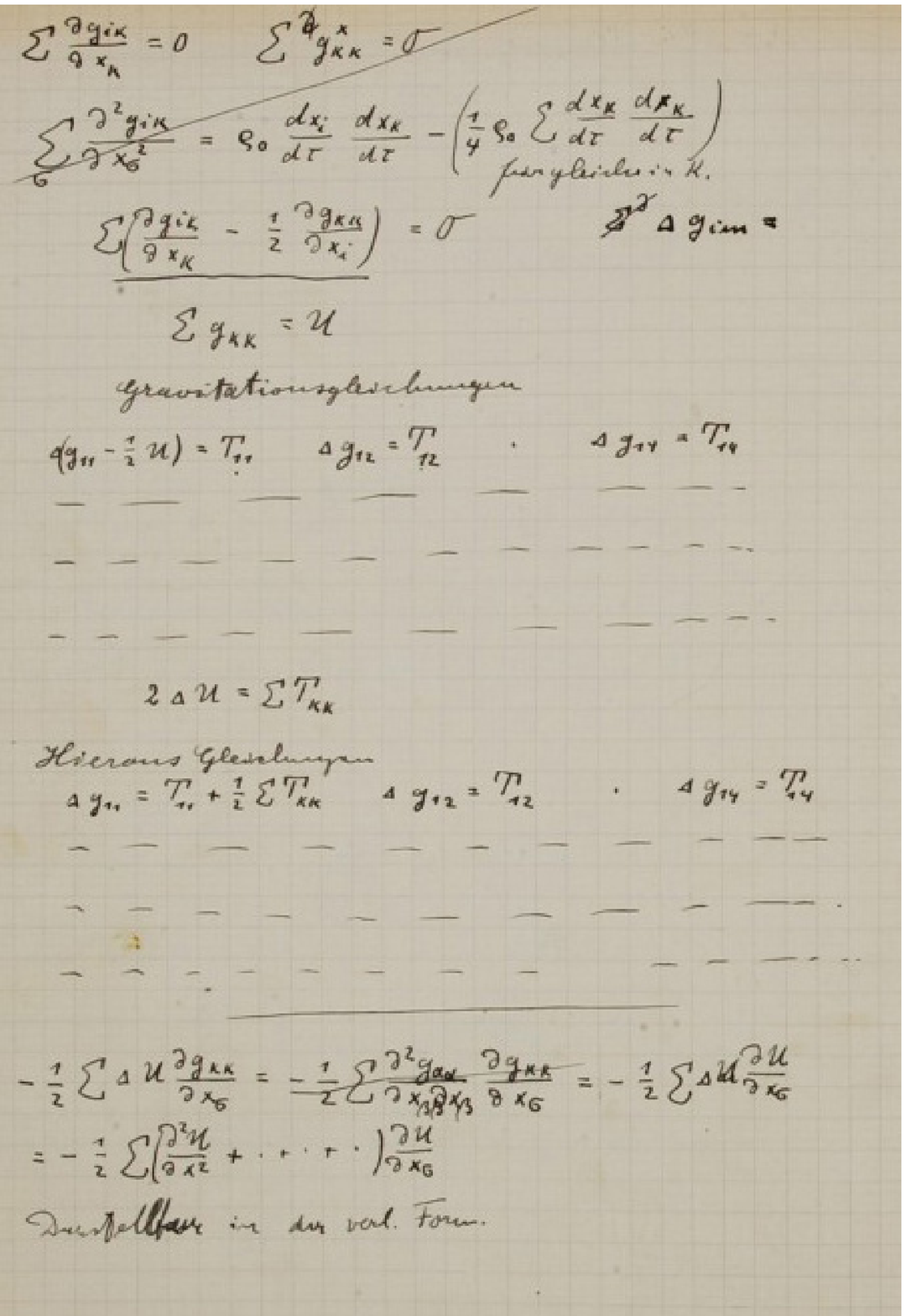}
\caption{Einstein's so-called ``Zurich Notebook'' contains research
notes on a relativistic theory of gravitation from the period of ca.\
August 1912 to spring 1913. This page shows ``gravitational
equations'' (``Gravitationsgleichungen'') that are equivalent to the
linearized form of the gravitational field equations of general
relativity published by Einstein in late 1915. For a facsimile of this
page, see Doc.~3-006, image 20, on {\it Einstein Archives Online} at
http://www.alberteinstein.info; for an annotated transcription, see
\cite[Doc.~10, esp. pp.~247f.]{CPAE4} \copyright~The Hebrew University
of Jerusalem, Albert Einstein Archives.}
\end{center}
\end{figure}
The very same equations were nevertheless published as the final field
equations of gravitation in November 1915 \cite[Doc.~25]{CPAE6}, and
this publication marks the breakthrough to the final theory of general
relativity, as we still accept it today. The reconstruction of this
particular research notebook thus offers unparalleled insights into
the considerations of a creative scientific mind that were eventually
crowned with remarkable success. A two-volume monograph on the genesis
of general relativity that gives an account of Einstein's search based
on a meticulous reconstruction of these notes is currently being
prepared for publication \cite{Rennetal2005}. It will include a
line-by-line commentary together with a facsimile and transcription of
the complete notebook.

In contrast to Einstein's early scientific achievements and especially
his special theory of relativity, his later path towards a general
theory of relativity is rather well documented by publications,
correspondence, as well as research notes. Among these documents the
Zurich Notebook is perhaps the most elucidating but there are
more research notes pertinent to this period. Another set of research
manuscript pages that bears testimony to Einstein's efforts of this
period presented another editorial challenge to the editors of Volume
4 of the series. It is a document that consists of 51 manuscript pages
plus additional text and calculations that were noted on a letter that
Einstein received. This set of research notes deals with the problem
of calculating the perihelion advance of Mercury on the basis of the
Einstein-Grossmann {\it Entwurf}-theory of gravitation of summer
1913. 

An anomalous secular advance of Mercury's perihelion had been well
established around the turn of the century, and since it could not
easily be explained in the framework of Newtonian gravitation theory,
this anomaly presented a touchstone for any alternative theory of
gravitation. After settling on the field equations of his preliminary
{\it Entwurf}-theory, Einstein set out to compute this problem together
with his friend Michele Besso. The manuscript documents their
collaboration. About half of the pages are written in Einstein's hand,
the other half in Besso's hand. The manuscript consists of loose
sheets and a determination of the proper sequence of the sheets
presented a major problem of the editorial work. But the coherence of
the notes made an almost complete reconstruction of the calculations
possible and on the basis of this reconstruction a likely original
sequence of the notes could well be established.

The full manuscript was presented in conformal transcription in Volume
4 \cite[Doc.~14]{CPAE4}. As in the case of the Zurich Notebook, the
document was prefaced by an editorial note elaborating specifically on
its context and content, and the pages were also annotated with
footnotes rather than endnotes. In contrast to the Zurich Notebook the
editors decided to complement the edition by a full facsimile of the
manuscript in an appendix.

Again, this particular research manuscript was subject of intensive
historical research. Its analysis helped explain why Einstein was very
quick in computing the correct value of the perihelion precession two
years later on the basis of his final field equations
\cite{EarmanJanssen1993}. This question is an important aspect of the
history of general relativity, since the successful explanation of the
Mercury perihelion anomaly sealed Einstein's breakthrough to general
relativity in November 1915 and played an important role in convincing
the physics community to accept the general theory of relativity.

No serious editorial challenges of this caliber were encountered for
the other two {\it Writings} volumes that have already been published. Both
volumes 6 and 7 also differ from earlier {\it Writings} volumes in that they
present unpublished manuscripts by Einstein that are of a
non-scientific nature, e.g.\ political statements. In addition to
Einstein's published papers, both volumes contain lecture notes, draft
manuscripts intended for publications, and expert opinions. Volume 7
presents some small research manuscripts in an appendix
\cite[App.~A]{CPAE7} and \cite[App.~B]{CPAE7}.  Here five pages of
calculations and graphical integration are reproduced in facsimile
without further commentary, since they are clearly related to a
published paper that is reprinted in the volume (Doc.~56). The five pages
are identified as coming from different sources and are presented as
one document on the basis of their contentual relation to the published
paper. The other manuscript appendix is a transcription of the first
part of an autograph manuscript by Einstein's collaborator Wander J.\
de Haas that de Haas attributes to Einstein as author together with a
transcription of a page of calculations for this part in Einstein's
hand.

\subsection{``Back-of-the-Envelope calculations''}

Another category of research manuscripts that are encountered
frequently in the Einstein Papers are small pieces of calculations or
notes scribbled on the verso of letters, on the back of envelopes,
etc. To the extent that these notes can be identified as being related
to the main document they are either included in the edited text or
transcribed, described, or discussed at the appropriate places in the
annotation. In many cases, however, the notes defy a clear
identification other than that they do not show any apparent relation to
the main document. These research notes pose a problem particularly
for the {\it Correspondence} volumes.

An example is given in Fig.~3. 
\begin{figure}
\begin{center}
\includegraphics{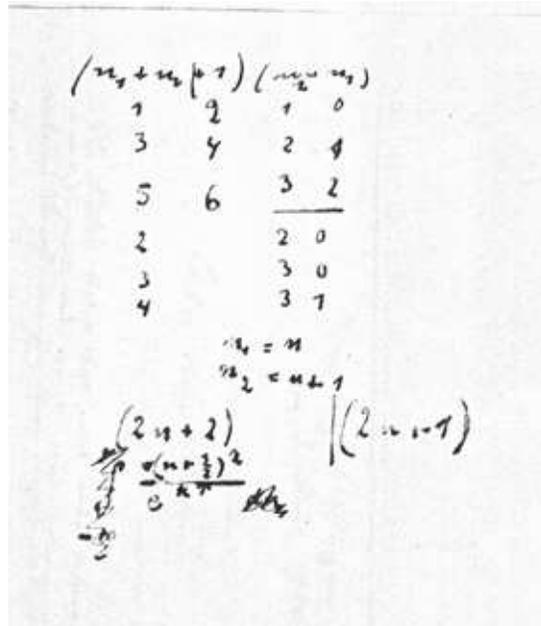}
\caption{Calculations on the verso of a letter by Anton Lampa to
Einstein, 1 February 1920 \cite[Doc.~291]{CPAE9}. Due to their disparate
and disconnected nature these notes were only described in the
descriptive note to the edited document. \copyright~The Hebrew
University of Jerusalem, Albert Einstein Archives.}
\end{center}
\end{figure}
The calculations are found on the verso of a letter by Anton Lampa,
formerly a colleague of Einstein's as professor at the German
University of Prague, who had recently accepted a position as Director
of Public Education at the School Department in Vienna. The letter
discusses a matter of appointment policy for a vacant chair in
Vienna. The calculations on the back are only mentioned in the
descriptive note which reads:
\begin{quote}
On the verso of the document, Einstein calculated expressions
involving quantum numbers $n_1=n$ and $n_2=n+1$ for a few values of
$n$, possibly related to Bjerrum's explanation of rotational spectra
(see Doc.~335 and its note 9). \cite[p.~397]{CPAE9}.
\end{quote}
The calculations are too short to be identified unambiguously and they
are too short to be published separately either as a document of its
own or in an appendix. But the cross reference to another document in
the volume indicates that the editors saw a possible connection to a
scientific problem that plays a role elsewhere in Einstein's
correspondence. Clearly such description of material that is omitted
from the edition represents a compromise between the wish to present
as complete an edition as possible and to present only documents that
have been identified with respect to content and
dating. It remains a judgment call for the editors to decide on the
level of detail to use in describing this material.

\subsection{The Berlin and Princeton Manuscripts on Unified Field Theory}

Presumably, many research manuscripts extant in the Einstein Archives
will pose no significant new problems from an editorial point of view,
exciting and illuminating as they might be for their content. There
is, however, a significant amount of material that does represent a
challenge for the editors of future volumes of the CPAE series: a
batch of some 1750 manuscript pages with calculations dealing mostly
with Einstein's search for a unified theory. Einstein began
thinking about the problem of a unified field theory of gravitation
and electromagnetism soon after the completion of the general theory
of relativity, and his first publications on this topic date from
1919. The problem remained on his mind through the next three-and-a-half 
decades until his death in 1955 \cite{Sauer2004}. Some fifty
publications from this period attest to his sustained interest in this
problem.

The manuscript pages are not part of the original Dukas
collection. According to anecdotal information, the batch of
manuscript pages, surprisingly, turned up behind a filing cabinet when
the archival offices used by the late Helen Dukas at the Institute for
Advanced Study in Princeton were cleared up in preparation of sending
the Einstein Estate to Jerusalem. The batch was then added to the
Einstein Archives and, after shipping to Jerusalem, it was microfilmed
into reels 62 and 63 and accordingly accessioned under archival call
numbers 62-001 through 63-416.

The manuscripts are undated pages, for the most part single sheets;
see Fig.~4 for an example. From the later years (reel 63), the paper
of these sheets appear to have been taken from loosely bound
typewriter paper tablets, many of them of the same kind. Some of the
sheets were bound together by means of paper clips. In some rare
cases, the versos of letters or other paper was used. The manuscripts
almost exclusively contain research calculations. They were made for
Einstein's own use, i.e.\ they very rarely contain explanatory remarks
about the context or meaning of the calculations. In some rare cases,
more elaborate parts of drafts of manuscripts are found in the
set. Some pages contain calculations and comments that are not in
Einstein's hand.
\begin{figure}
\begin{center}
\includegraphics[scale=.6]{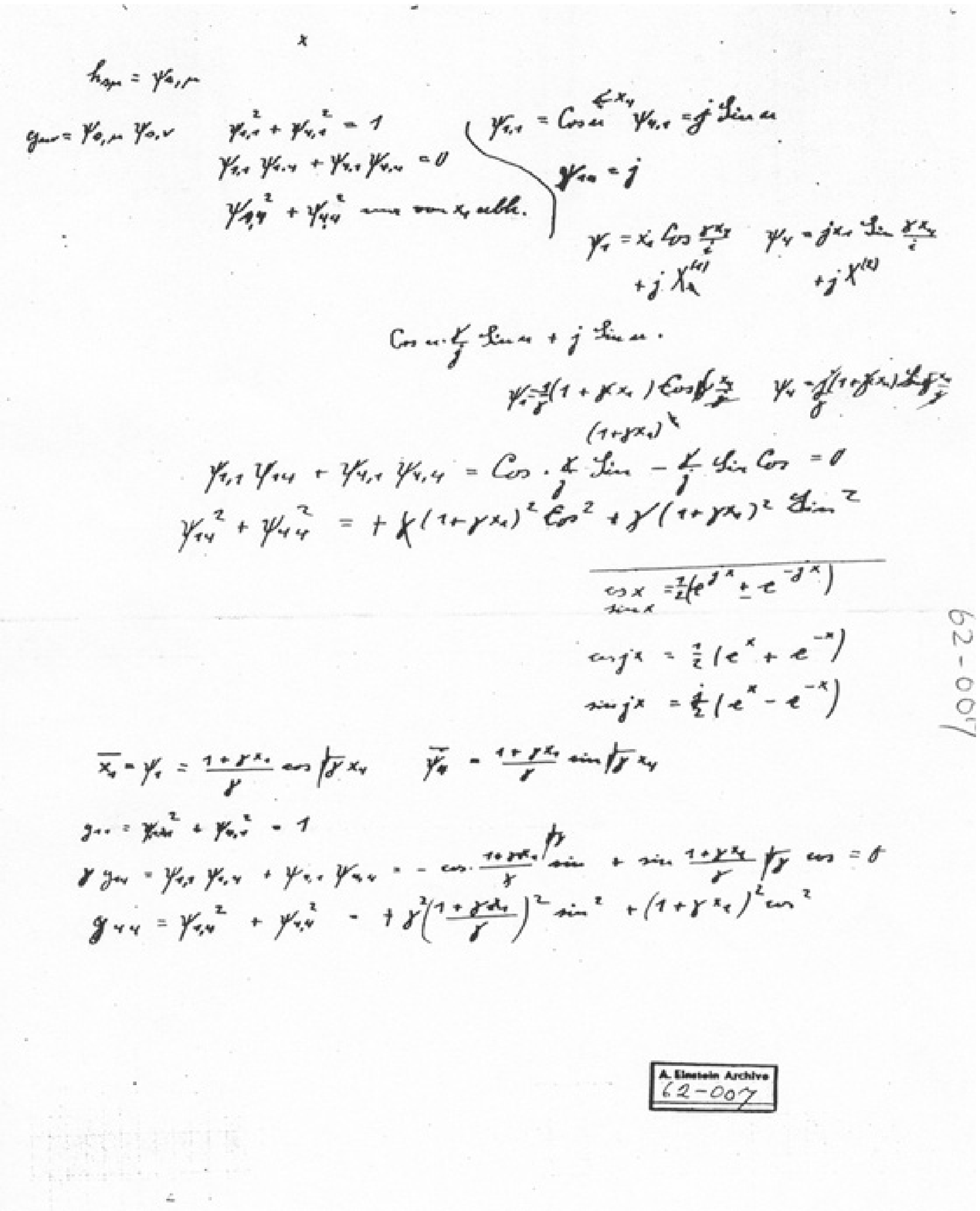}
\caption{A page of Einstein's research notes on unified field theory
(Call Nr.~62-007). This particular page can be identified as being
related to a publication from 1929. There are more than 1700 undated
pages of this kind in the Einstein Archives. This batch of research
manuscripts presents a major challenge for future volumes of the {\it
Collected Papers of Albert Einstein}.  \copyright~The Hebrew
University of Jerusalem, Albert Einstein Archives.}
\end{center}
\end{figure}

The material poses a problem for the editorial project specifically in
all of the aspects discussed so far, i.e.\ dating, coherence, and
chronological as well as logical sequence of the pages. Two aspects of
the material add to its being an editorial challenge. For one,
the sheer amount of the material precludes any detailed or
in-depth analysis without some global assessment of what is
to be expected. Second, in contrast to Einstein's manuscripts from his
earlier years, these manuscript do not document a productive
creativity that eventually led to a successful theory. Hence, to this
date historical interest in Einstein's investigations into a unified
field theory has been rather limited, some important exceptions
notwithstanding.\footnote{For references to secondary literature on the
history of unified field theory, see \cite{Goenner2004} and
\cite{Sauer2004}.}

With respect to dating, a superficial survey of the material showed at
least one example where calculations were found on an envelope that
carries a poststamp of 1919 (Call Nos.~62-052 and 62-053). Since this
implies a terminus a quo that would perhaps render some of the
material relevant for volumes of the CPAE series that are presently
under preparation, a more detailed investigation of the whole set of
pages was called for.

In order to better be able to deal with the material, copies extant in
the duplicate Archives at the Einstein Papers Project were scanned in
as images of low-resolution black-and-white quality. The pages were
then looked at individually using the scanned version, and a
multipurpose database application was used to establish a page-by-page
catalog of the manuscript pages. The information recorded in this
database pertained primarily to any legible word or phrase that could
be deciphered on the page. In the example of the page reproduced in
Fig.~4, the only legible phrase in this sense is in the fourth line
and reads: ``nur von \$x\_1\$ abh.'' The mathematical formulae and
variables were generally completely ignored. In some cases, where it
seemed useful to also enter a mathematical expression or formula, that
expression or formula was transcribed using standard
LaTeX-markup.\footnote{The LaTeX-transcription provides a
representational description not a contentual one. It also does not
allow, in general, to search for specific mathematical expressions
since the same mathematical expression can be represented in several
ways using LaTeX-markup.  As a simple example, the LaTeX-markups
\$x\_1\$ and \$x\_\{1\}\$ are representationally equivalent.}  Any
obvious pecularities of the manuscript pages, such as specifics of the
paper or ink, or entries in a different hand, etc. were also added to
the records. Most importantly, any information that is relevant for
dating a particular page was noted. Such information comes from the
use of dated sheets of paper, e.g.\ verso of letters or draft of
letters or use of hotel stationery; from an unambiguous identification
of content of the material; from the appearance of a different
handwriting or the mention of a collaborator of Einstein's; from the
mention of identification of specific literature; and similar hints.

The descriptive database of the set of manuscript pages in conjunction
with scanned images of the manuscript pages represents a working tool
for further analysis of the manuscripts. Thus, unidentified copies of
certain manuscript pages extant in the supplementary archives could in
part be identified with known ones by searching for the legible word
or phrase content. Also, generic concordance tools of computational
linguistics are applicable to create word indices for the manuscript
pages. Despite the preliminary character of the database, some facts
about the manuscript pages could already be established.

Putting together a survey of all dated or datable pages, it appears
highly likely that none of the material dates any earlier than
1928. This is probably also true for the single instance of Call No
62-052 which remains undated but may well be much later than the post
stamp of 1919. A very strong indication for taking 1928 as the
earliest date for the whole manuscript set is given by the fact that,
in general, independent hints for dating of different pages tend to
corroborate each other for pages that are close in the sequence of
sheets, and all those hints point to dates later than 1928. It also
seems that the sequence of sheets in which they are preserved roughly
reflects a chronological order, although exceptions are very well
posible and in some cases even likely. Hence, the manuscript pages are
removed from the immediate concerns of the editors of the CPAE
series. Nevertheless, they continue to represent a major editorial
challenge for future volumes.

\section{Concluding Remarks}

In order to be published as an individual document in a {\it Writings}
volumes of the CPAE series, research manuscripts need to meet at least
two criteria. They need to be dated with sufficient accuracy and they
need to be sufficiently coherent so as to be identified as one
document. In the case of the Scratch Notebook, 1910--1914?, the
coherence was primarily suggested by the fact that the notes were
entered in a physically bound notebook. But neither could the notes be
dated accurately, nor could a unity of content be established. Hence,
the notebook was reproduced in its entirety in an appendix. In the
case of the Zurich Notebook, only a part of the entries in the bound
notebook could be identified as a sufficiently coherent and datable
set of research notes, and that set of pages was included
as an edited document. The Einstein-Besso manuscript could be dated
and identified as a coherent set of notes through a reconstruction of
its content. Many small ``back-of-the-envelope'' calculations cannot be
edited as texts because their meaning and signficance cannot be
identified. Since they are part of another document, they
are in many instances at least described in the descriptive note of
the respective document.

With our present understanding of Einstein's Berlin and Princeton
research manuscripts on unified field theory, a considerable amount of
scientific manuscripts left by Einstein defies the procedures of the
documentary edition of the CPAE. Neither can we already date the
material with sufficient accuracy, nor do we have a sufficient
understanding of their coherence and of their chronological or logical
sequence.

Does the launch of the {\it Einstein Archives Online} website change
the situation by providing other ways of publishing important
documents? High-quality images of both the Scratch Notebook as well as
of the full Zurich Notebook are, in fact, now available
online. The Einstein-Besso manuscript, however, is not included on the
present website, since the original is not part of the holdings of the
Albert Einstein Archives.\footnote{The manuscript is in private
possession. It was auctioned through Christies in October 2002.} And none
of the small ``back-of-an-envelope'' calculations are presently
available on the site since it does not include any correspondence. It
would indeed be a natural expansion of the present website to include
at least facsimiles of those letters that are not included in the
respective {\it Correspondence} volumes. It is also conceivable that
facsimile images of Einstein's Berlin and Princeton manuscripts might
be added to the document content of the {\it Einstein Archives Online}
website before it is published in the documentary series.

Nevertheless, a mere publication of the manuscript images online does
not represent a proper documentary editing of this important
manuscript material. The database content that goes with the
facsimiles of the documents on the website is far less explicative and
not as reliable as the carefully established information in the
documentary edition. More importantly, no transcription, no
translation, and none of the editorial apparatus of the documentary
edition is included. I hope to have shown that a successful response
to the editorial challenge posed by Einstein's research material goes
hand in hand with a scholarly investigation into its content and
context. Only by trying to understand the content and context of the
manuscript will we be able to satisfactorily edit those manuscripts, be it
in conventional large book format or electronically. We may also hope
to get further exciting insights into the workings of a highly
creative mind.

\end{document}